\documentclass[conference]{IEEEtran} 
\usepackage[dvips]{graphicx}
\usepackage{subcaption}
\usepackage{epsf}
\usepackage{epsfig}
\usepackage[noadjust]{cite}
\usepackage{amsmath}
\usepackage{url}
\usepackage[export]{adjustbox}
\usepackage{soul}
\usepackage{algorithm}
\usepackage[noend]{algpseudocode}
\usepackage{color}
\usepackage{tree-dvips}
\usepackage{float}
\usepackage{multirow}
\usepackage{comment}
\usepackage{soul}
\usepackage{booktabs}
\usepackage[dvipsnames]{xcolor}
\usepackage{siunitx}
\usepackage{footnote}
\usepackage{threeparttable}
\usepackage{pifont}
\usepackage[T1]{fontenc}
\usepackage[algo2e]{algorithm2e}

\def\BibTeX{{\rm B\kern-.05em{\sc i\kern-.025em b}\kern-.08em
    T\kern-.1667em\lower.7ex\hbox{E}\kern-.125emX}}

\begin{document}

\onecolumn 

{\LARGE IEEE Copyright Notice} \\

\copyright 2019 IEEE. Personal use of this material is permitted. Permission from IEEE must be obtained for all other uses, in any current or future media, including reprinting/republishing this material for advertising or promotional purposes, creating
new collective works, for resale or redistribution to servers or lists, or reuse of any copyrighted component of this work in other works. \\

{\large Accepted to be Published in: Proceedings of the 2019 5th IEEE International Symposium on Smart Electronic Systems (IEEE-iSES), Dec. 16-18, 2019, Rourkela, India.}

\twocolumn

\title{DLockout: A Design Lockout Technique for Key Obfuscated RTL IP Designs \vspace*{-1ex} } 

\author{\IEEEauthorblockN{Sheikh Ariful Islam, Love Kumar Sah and Srinivas Katkoori}
\IEEEauthorblockA{Department of Computer Science and Engineering\\
University of South Florida \\
Tampa, FL 33620\\
Email: \{sheikhariful, lsah, katkoori\}@mail.usf.edu}
\vspace*{-6ex}}



\maketitle

\begin{abstract}
Intellectual Property (IP) infringement including piracy and over production have emerged as significant threats in the semiconductor supply chain. Key based obfuscation techniques (i.e., logic locking) are widely applied to secure legacy IP from such attacks. However, the fundamental question remains open whether an attacker is allowed an exponential amount of time to seek correct key or could it be useful to lock out the design in a non-destructive manner after several incorrect attempts. In this paper, we address this question with a robust design lockout technique. Specifically, we perform comparisons on obfuscation logic output that reflects the condition (correct or incorrect) of the applied key without changing the system behaviour. The proposed approach, when combined with key obfuscation (logic locking) technique, increases the difficulty of reverse engineering key obfuscated RTL module.  We provide security evaluation of DLockout against three common side channel attacks followed by a quantitative assessment of the resilience.  We conducted a set of experiments on four datapath intensive IPs and one crypto core for three different key lengths (32-, 64-, and 128-bit) under typical design corner. On average, DLockout incurs negligible area, power, and delay overheads.

\end{abstract}

\section{Introduction}
\label{sec:intro}

In recent decades, horizontal IC business model and vertical disintegration of the design have proven themselves to manufacturing and testing of fabless design houses' IP/IC in foreign foundries \cite{6069789}.  This trend becomes attractive for a system integrator to integrate Commercial Off the Shelf (COTS) components to meet a strict requirement for time to market.
In the heart of this design ecosystem, original IP owners face several security challenges. Frequent IP handover in the supply chain could pose the IP to be vulnerable to unauthorized duplication and piracy. Reverse engineering is commonly employed to execute the variants of IP theft. To reduce the risk of IP theft, the promise of obfuscation transforms the original IP into an equivalent design with  a greater barrier to uncover  functional semantics without the correct key.


The use of hardware obfuscation approaches \cite{4681649,Desai:2013:IOA:2459976.2459985} in recent years is becoming common practice to protect the legacy RTL IP. These approaches perform transformations to the original FSM by embedding additional states and depending on key value, they  control \textit{modes of operation}. At the same time, state-of-the-art obfuscation methods have been found to protect only RTL Hardware Description Language and lacks flexibility in securing both the datapath and controller of an RTL design. Furthermore, existing obfuscation solutions at RT-level incur substantial performance overhead while building the security into hardware.   Despite the major objective of key-based obfuscation or logic locking, one important question remains unanswered: is the brute-force attempt something that can be complemented with early locking out the design for finite but incorrect attempts?

Motivated by the software IP licensing scheme,  we present a low-cost lockable obfuscation framework, DLockout, for hardware IP. In software  regime, a software owner favors a user (attacker) to a finite number of attempts for legal use of the software. After a finite number of incorrect tries, the user is requested to provide another form of verification to regain access to locked software IP. We do this in obfuscated RTL IP by embedding comparators (XORs) to the obfuscation logic (MUXes) in non-critical paths. Following that, we introduce a counter to be compared with a preset threshold. This threshold determines to what extent (number of attempts) the (in)correct key can be applied. For each incorrect key retrieval attempt, the counter value is incremented and when it reaches the threshold, obfuscated IP is locked out.  We then introduce a checker FSM in obfuscated datapath that signals the controller during the lockout to enter into a {\em blackhole} state.  It  ensures that all  accesses to the design by the legitimate users are valid as long as the correct key is applied, thus maintaining the design for security. The design lockout approach runs together with the original functionality of the design and requires minimal changes in the obfuscated RTL IP. To the best of our knowledge, DLockout  is the first comprehensive technique that enhances traditional key hardware obfuscation with lockout mechanism. In summary, the novelty of  DLockout includes: (i) no storage of the key within the obfuscated design;  (ii) stealthy key propagation and comparison through non-critical path(s);  and (iii) minimal modifications to an existing obfuscated RTL design with lightweight, low-overhead comparators and checker FSM.




The remainder of the paper is organized as follows. Section \ref{sec:backgnd}
reviews the necessary background and previous RTL obfuscation schemes. Section \ref{sec:attack} presents the attack model followed by proposed DLockout architecture in Section \ref{sec:Dlock}. We provide the security evaluations and experimental results in Section \ref{sec:sec_eval} and \ref{sec:result} respectively. Section \ref{sec:conclusion} concludes the paper.

\section{Background and Related work}
\label{sec:backgnd}




In this section, we review state-of-the-art literature on key-based hardware obfuscation and obfuscating transformations at RTL against IP piracy. Mode-based RTL obfuscation has been studied in \cite{Chakraborty:2010:RHI:1729649.1730613}. The technique works by extending the bit-length of host registers for appropriate mode selection with moderate overhead. The authors in \cite {5224963} constructed the Control Flow Graph (CFG) of the RTL description from  gate-level designs. In this obfuscation model, extra decision nodes depending on boolean computations of state elements are inserted.  \textit{ObfusFlow} \cite{4681649} performs the  XOR operation to a subset of internal nodes with an additional FSM.  The Boosted Finite State Machine (BFSM) \cite{207766} hardens the original FSM by additional states. The technique powers up the design with additional states such that a random unique block is entered before going into working mode else \textit{blackhole state} mode is entered. The authors in \cite{8357321} utilized
mobility of input operands during HLS to increase stealthiness of RTL obfuscation. Although the technique introduces the obfuscation during the early design, it does not include the abuse case when an attacker may want to retrieve the key for a finite number of times. The published RTL obfuscation works did not consider key management block built in to prevent unauthorized execution of RTL IP. We distinguish our work as follows: (a) the key propagation is as stealthily as possible and locks out the design for incorrect key when certain number of attempts are made and (b) we provide the security evaluation of an RTL design against state-of-the-art side channel analysis.

\begin{figure}[h]
  \centering
  \vspace{-4ex}
  \includegraphics[width=\columnwidth]{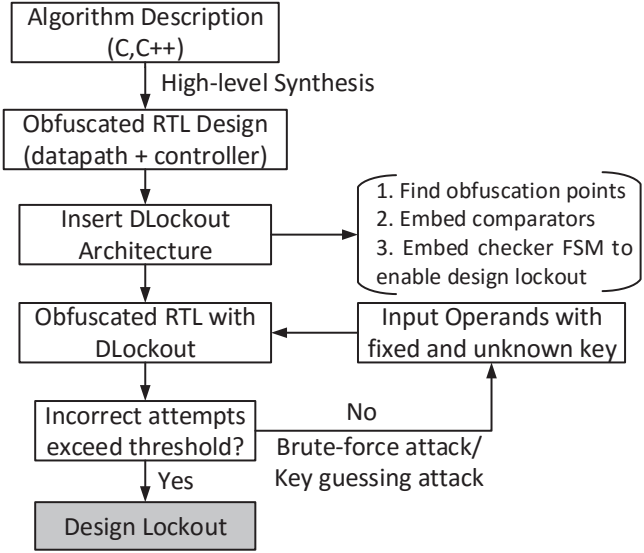} 
  \vspace{-2ex}
  \caption{DLockout integration in key obfuscated RTL design}
  \label{fig:dlock}
  \vspace{-4ex}
\end{figure}

\section{Threat Model}
\label{sec:attack}


To reverse engineer the correct key, we assume an attacker has access to a  logic-level locked but flattened netlist and an oracle black-box IC. We also assume that the design cannot be subjected to sophisticated micro-probing attack (e.g. circuit edit \cite{7984893}). The published works in \cite{6241494,7428086} made the similar assumptions for the threat model where the deobfuscation intent was effective either on fully combinational  or sequential design at the gate-level. In addition, we assume that the attacker's objective is to reveal the key to enabling the true functionality of the obfuscated RTL design. We also assume that the attacker does not have access to internal nets except primary input(s) and output(s). Under these assumptions, the adversary can apply any input sequence, observe the output, and analyze the input-output behavior.

\begin{figure*}[h]
  \centering
  \includegraphics[width=\textwidth]{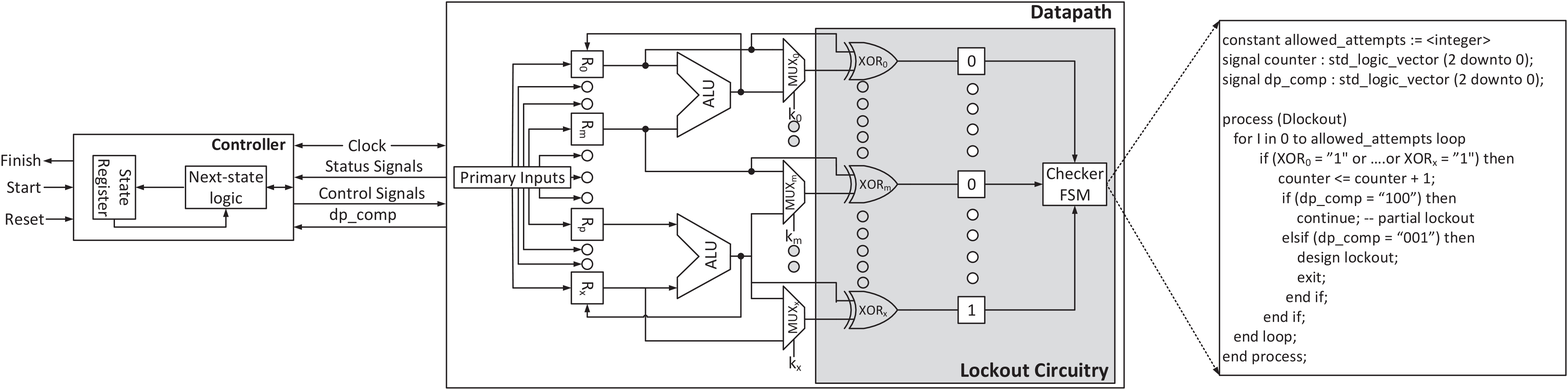} 
  \vspace*{-3ex}
  \caption{Glushkovian model of key obfuscated RTL design with design lockout circuitry}
  \label{fig:glas_arch}
  \vspace*{-3ex}
\end{figure*}

\section{Proposed DLockout Architecture}
\label{sec:Dlock}

Our proposed technique, DLockout, in Fig. \ref{fig:dlock} defends against counterfeiting and fights against reverse engineering aiming at obtaining design data of IP. We ensure this by by embedding comparators at the output of obfuscation logic after regular key obfuscation is performed. As shown, it contains three major steps for key verification. First, given a key obfuscated RTL IP, the designer would incorporate DLockout using the method proposed in Sections \ref{sec:dlock_dp} and \ref{sec:dlock_ctrl}. Secondly, an attacker (user) during the post-synthesis stage of RTL in the supply chain would apply regular input(s) and key bit(s) of a particular length. As the extraction of the key  is the most frequent target in any obfuscation scheme, one can apply brute-force  or intelligent key inference techniques. A check between the allowed attempts and the number of times (in)correct key applied will enable or disable DLockout. Finally, when unsuccessful key extraction trials end, the design is self-locked out permanently in a non-destructive manner.

\subsection{Design Lockout in RTL Datapath}
\label{sec:dlock_dp}

DLockout relies on the observation that obfuscated RTL IP is available after the designer performs scheduling, allocation, and binding according to cost-speed trade-off during HLS \cite{8357321, 8356053}. we follow the works in \cite{8357321} to determine the place of insertion of MUXes once the multiplexer based key at suitable obfuscation points in an RTL datapath are performed.  We then  annotate these obfuscating points with XORs to verify the applied key are correct for the successful execution of the design. These annotations are followed by a counter to ensure that the number of the times the incorrect key can be applied does not exceed the designer specified threshold, thus (semi)blocking the brute-force approach.  To ensure that an attacker would not utilize ``cold-reboot'' to reset counter value and be benefited from brute-force attempt, we assume the counter would be stored in non-volatile storage. We augment the datapath with a 3-bit, 2-input comparator. The annotated XORs' output determines the comparator output and introduces two variants (partial lockout and full lockout) of DLockout architecture. Even though it is possible for an attacker to find this regular structural pattern (MUX with XOR) inside an obfuscated RTL netlist, it is not possible to use this pattern to reveal the key.  Moreover, attackers cannot bypass the correct key enforcement as the existing obfuscation logic is used during both regular operation and key propagation for maximum flexibility and key interference during datapath synthesis.

\begin{figure*}
  \centering
  \includegraphics[width=\textwidth]{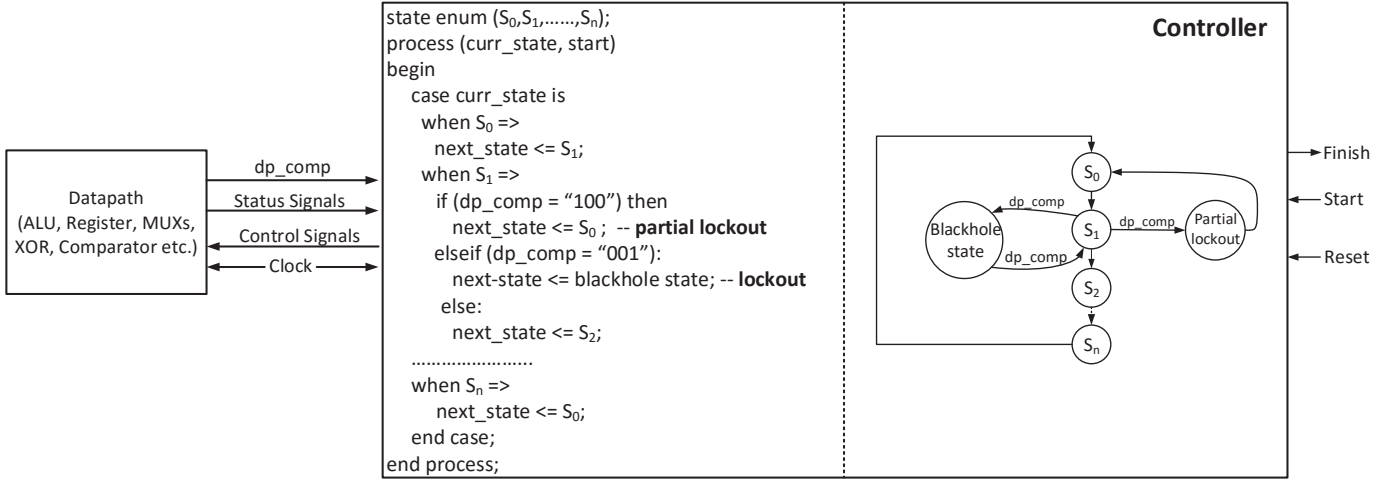} 
  \caption{RTL Controller with DLockout Architecture}
  \label{fig:ctrl}
  \vspace*{-4ex}
\end{figure*}

Fig. \ref{fig:glas_arch} shows the proposed modifications to Glushkovian model of an obfuscated RTL design. Each obfuscation logic in the datapath  is annotated with a 2-input XOR. The output (=0) of XOR  determines the correctness of key bit to a multiplexer selector input. 
The checker FSM would check all available XORs' output and increment the counter for any annotated XOR output signal being equal to `1'. When such number of `1's reach the threshold for incorrect attempts, DLockout will be active in place. Partial lockout is active when a wrong key is applied to the obfuscation logic for the  first time. Design\_lockout determines all allowed trials have been made. In both cases, an exception will be raised and datapath will send out the output of the checker FSM ({\tt{dp\_comp}}) to the controller.

\subsection{Design Lockout in RTL Controller}
\label{sec:dlock_ctrl}

We integrate DLockout architecture into the RTL controller as shown in Fig. \ref{fig:ctrl}. We modify parts of state transitions to make it difficult for the attacker to correctly reconstruct the state transitions. The next state of the current state will change dynamically depending on the key bit and  {\tt{dp\_comp}} signal from the obfuscated RTL datapath. So, the period in which the modified controller is the same as the original controller is during when the design starts execution (i.e. Reset is high). After the primary input being latched into available registers and  key checking at the first control step ($S1$), the next state could be the truly original state ($S2$) for correct key or it would revert  to initial state ($S0$) for partial lockout. In the design lockout phase, any true valid state is stripped from becoming the next state and the controller would enter into a permanent {\textit{blackhole}} state as introduced. This adds more confusion to the attacker.

\begin{figure*}[h]
\resizebox{\textwidth}{!}{
\begin{tabular}{cccc}
\includegraphics[width=1.1\textwidth,right]{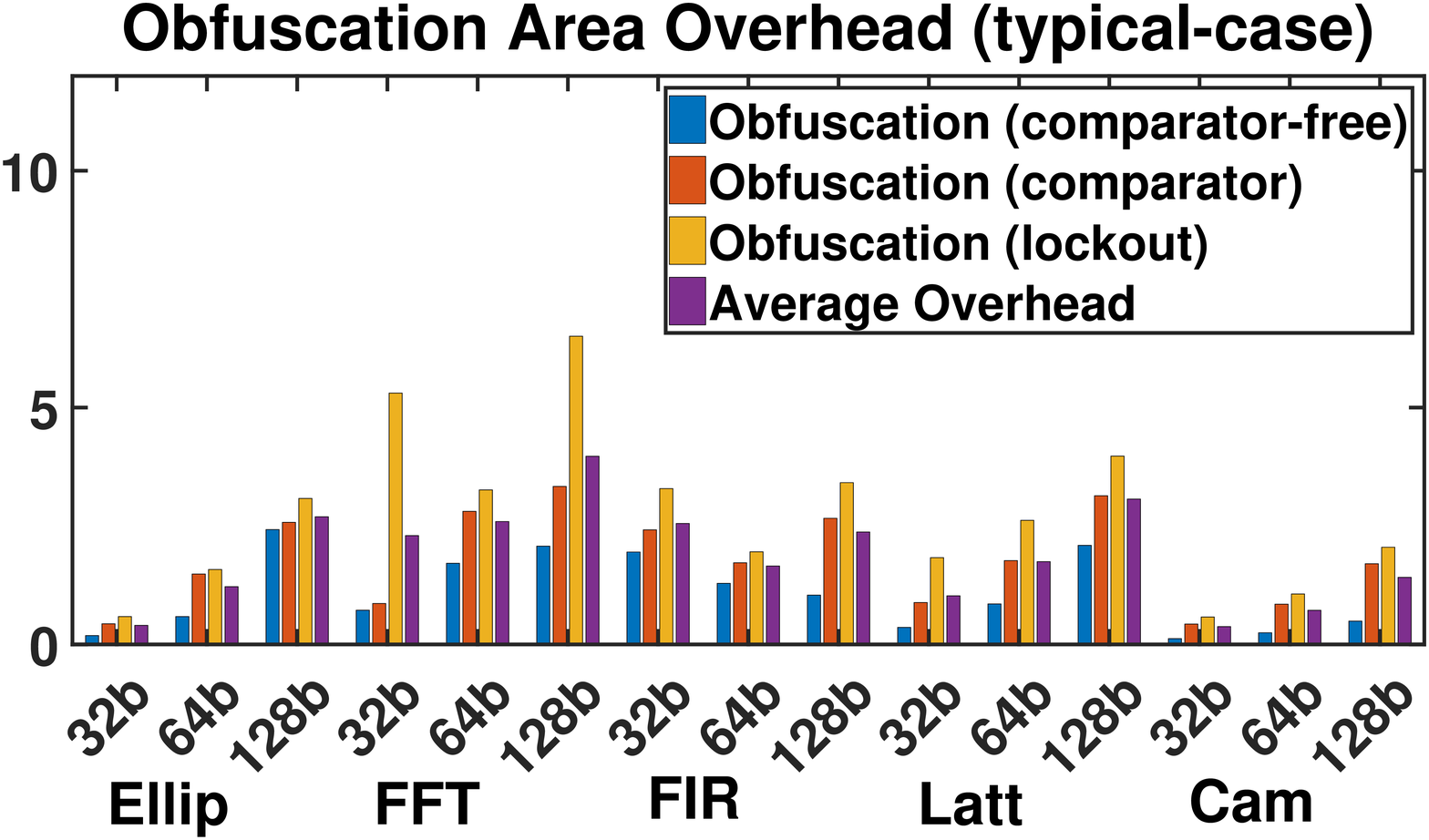} & 
\includegraphics[width=1.1\textwidth,right]{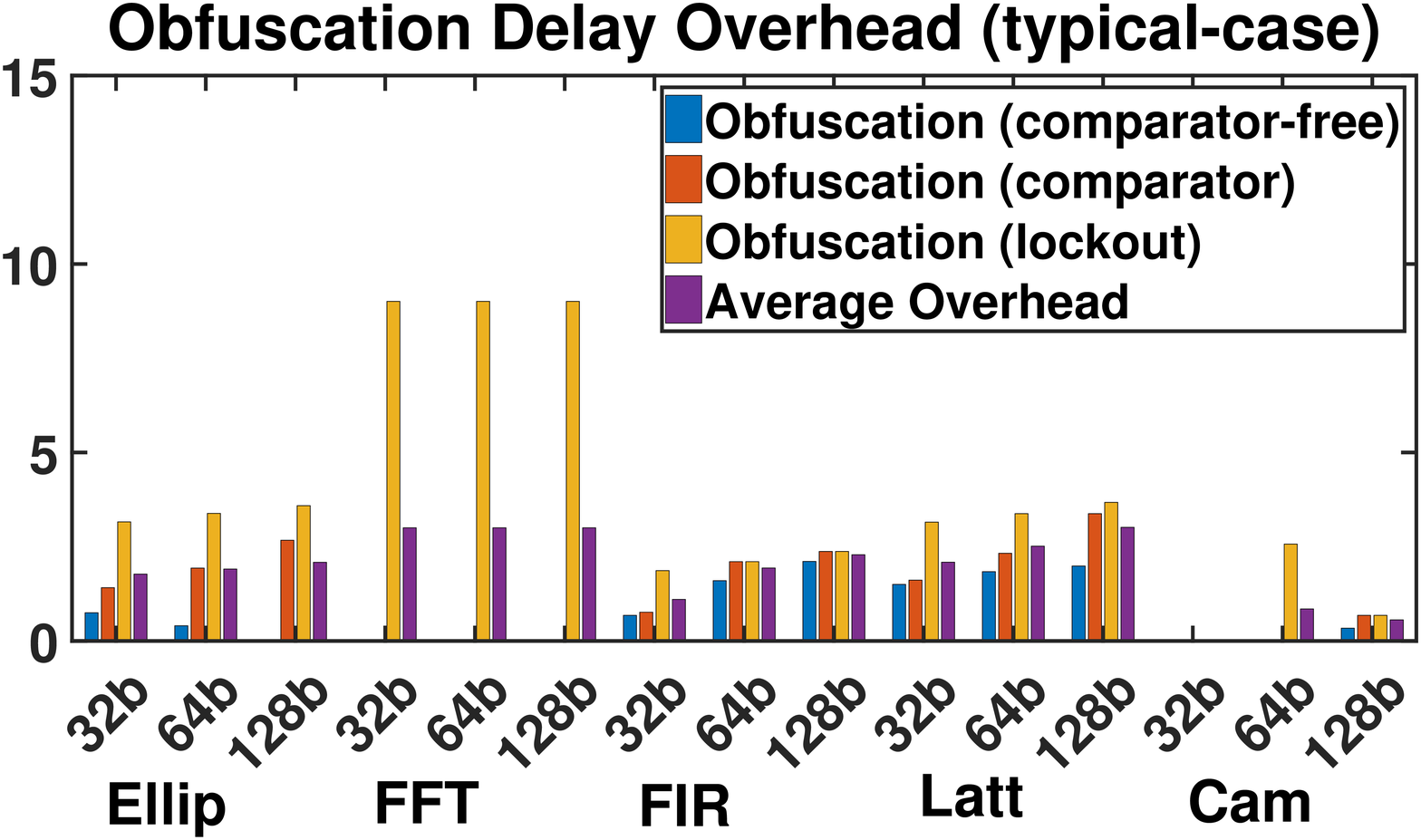} & 
\includegraphics[width=1.1\textwidth,right]{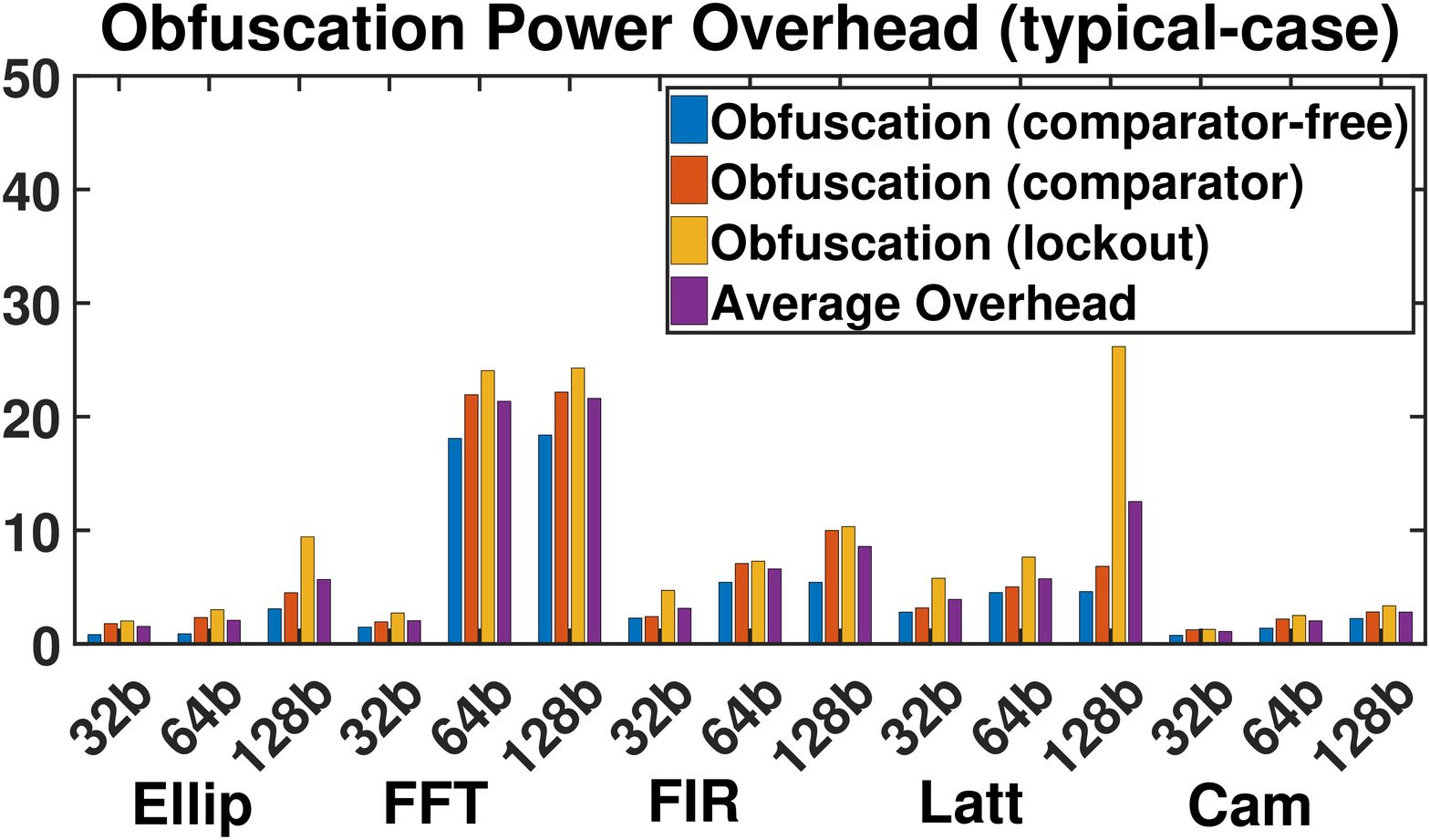} & 
\includegraphics[width=1.1\textwidth,right]{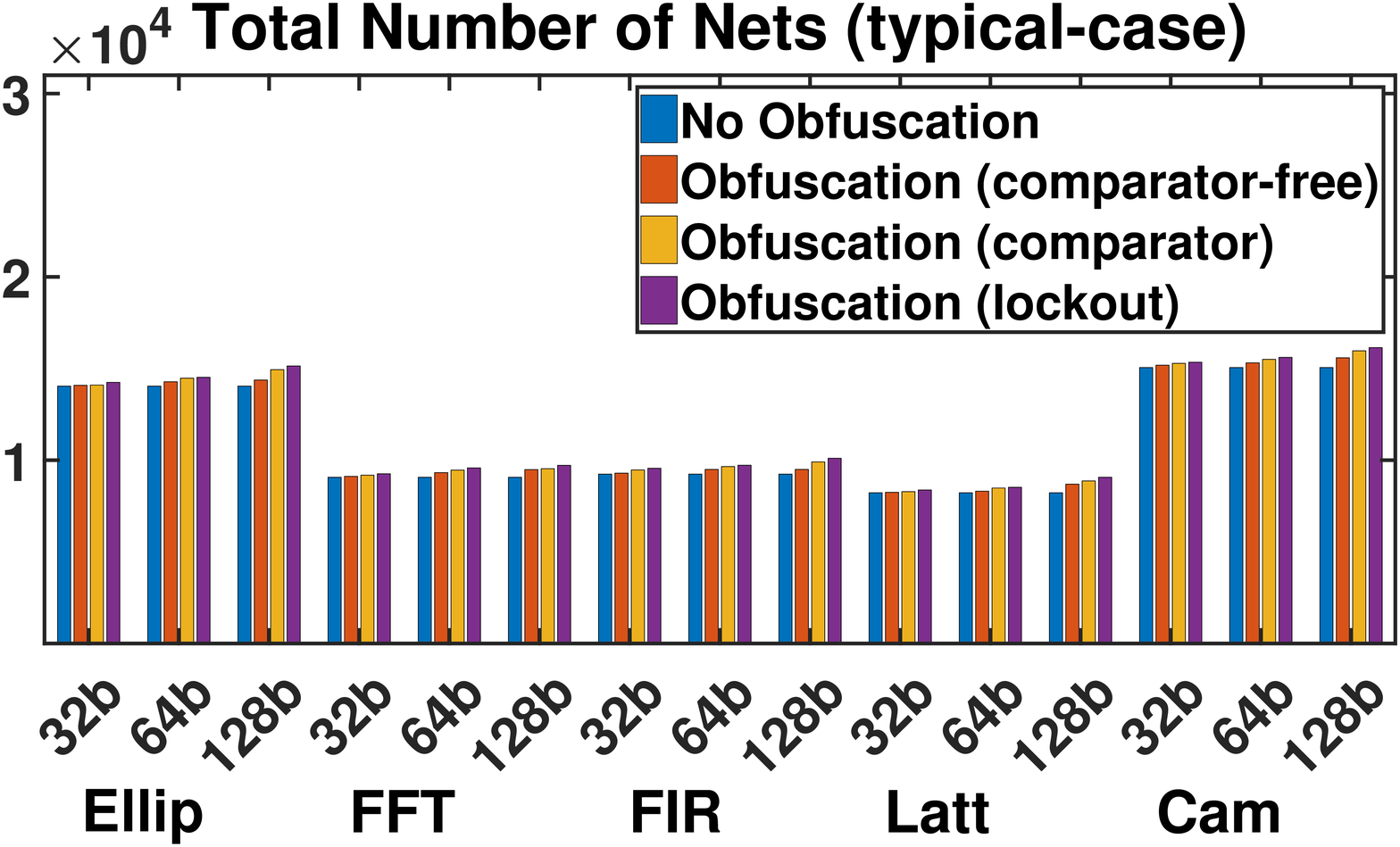} \\
{\Huge (a)} & {\Huge (b)} & {\Huge (c)} & {\Huge (d)}
\end{tabular}
}
\vspace*{-1ex}
\caption{Comparison of benchmark parameters for three key length in typical design corner}
\label{fig:singleplot}
\vspace*{-3ex}
\end{figure*}

\section{Security Evaluations}
\label{sec:sec_eval}

{\textbf{Power Analysis (PA) Attack}}: We provide a methodology to mount simulated Differential Power Analysis (DPA) attack on DLockout as follows. If the total number of switching bit is $p$ in a given $q$ bit MUX, we can measure the correlation coefficient,$r_{o}$, between input data and dynamic power consumption of an obfuscation logic. 
\vspace{-2ex}
\begin{equation}
\vspace{-1ex}
r_{0} \simeq  \sqrt{\frac{p}{q}}; p \leq q 
\end{equation}

Similarly, the relation between MTD (Measurements to Disclosure) and $r_{o}$ is given by:
\vspace{-2ex}
\begin{equation}
\vspace{-1ex}
MTD \propto  \frac{1}{r_{0}^{2}}
\label{mtd}
\end{equation}

If there are N control steps and M obfuscation logic are non-uniformly distributed across N, Eqn. \ref{mtd} will become:
\begin{equation}
\vspace{-1ex}
MTD_{1} \simeq \frac{MN}{r_{1}^{2}}*MTD_{0} \simeq \frac{MN}{r_{1}^{2}}*\frac{C}{r_{0}^{2}}
\end{equation}
where  C is the success rate dependent constant \cite{1580507} and $r_{1}$ is the correlation coefficient of the dynamic power consumption between MUX and other RTL components. To de-correlate power traces with the applied key, a random mask bit is XORed with the key bit to implement key-independent MUX output as shown  in Fig. \ref{fig:mask}. In other words, MUX output exhibits  the same probability distribution independent of key and mask bit. The  XOR output is also uncorrelated to individual key and mask bit as shown in Table \ref{table:mask_logic}. 


\begin{table}
	\begin{minipage}{0.49\columnwidth}
		\centering
		\includegraphics[width=45mm]{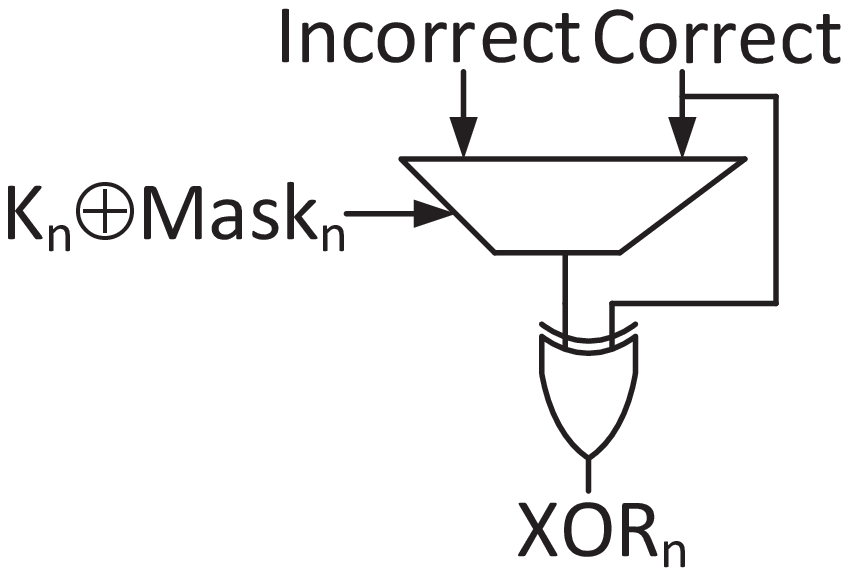}
		\captionof{figure}{Masked Obfuscation logic with comparator}
		\label{fig:mask}
	\end{minipage}
	\begin{minipage}{0.5\columnwidth}
		\caption{Truth Table of Masked Obfuscation logic}
		\label{table:mask_logic}
		\centering
		\begin{tabular}{lrrr}
			\toprule
			K$_{n}$          & Mask$_{n}$ & MUX$_{n}$ & XOR$_{n}$ \\
			\midrule
			0  & 0      &  Correct & 0\\
			0  & 1      & Incorrect & 1\\
		    1  & 0 &  Incorrect &  1\\
			1       &    1   &  Correct &  0\\
			\bottomrule
		\end{tabular}
	\end{minipage}\hfill
\vspace{-3ex}
\end{table}

\begin{table}
\centering
\caption{Operation of Error Detection Unit (EDU)}
\vspace{-1ex}
\label{table:edu}
\resizebox{\columnwidth}{!}{\begin{tabular}{lcccc}
\toprule
SAF at & Multiplexer & XOR Expected & EDU  & Comment \\ 
 XOR &  Output  & Output  & Output & \\ \midrule
0  & 0 (Correct) & 0 & 0 & Ineffective fault \& lockout-free \\
0  & 1 (Incorrect) & 1 & 1 & Fault detected \& lockout-free \\
1  & 0 (Correct) & 0 & 1 & Fault detected \& lockout \\
1  & 1 (Incorrect) & 1 & 0 & Ineffective fault \& lockout\\
\bottomrule
\end{tabular}}
\vspace{-5ex}
\end{table}

{\textbf{Fault Attack (FA)}}: In DLockout, the comparators' output exhibits the highest sensitivity to  deliberate fault injection. To detect  stuck-at-fault (SAF), we propose to incorporate an Error Detection Unit (EDU) in the datapath that checks the output of MUXs and XORs in Table \ref{table:edu}. Even with the incorrect key, the lockout architecture is susceptible to FA when the SAF at XOR is 0.  For example, if an attacker owns $n$ copies of the device where the number of attempts is $X$ for each copy, he can try out different keys on each one. If each device is locked with key bit of size $m$, still an attacker has to try 2$^{m}/(n*(X-1))$ trials for all copies. Here, during (X-1)$^{th}$ attempt, an attacker would apply fault technique as crossing X$^{th}$ trial would lead to permanent locking.

{\textbf{Key Extraction Probability}}:
If $n$ and $m$ denote the total number of obfuscation logic and the key size respectively, the probability of mapping a key from $2^m$ combinations to $n!$ permutations of MUXs is given by:
\begin{equation}
\vspace{-1.5ex}
    P(m,n) = \frac{1}{n!*2^m}
\end{equation}
If the total number of allowed attempt is $X$, the probability of guessing the correct key bit at $K^{th}$ attempt (K$\leq$ X):
\begin{equation}
\vspace{-1ex}
    f(K,X, P(m,n)) = {X \choose{K}}*P(m,n)* (1-P(m,n))
\end{equation}

\section{Experimental Results}
\label{sec:result}

We evaluate DLockout performance with the gate-level simulations for target clock period of 10ns and three different key length on four datapath intensive benchmarks (Elliptic, FFT, FIR, and Lattice) and one crypto core \mbox{(Camellia) \cite{opencores}}. 
Table \ref{tab:dpa_meas} lists the number of traces required to leak a single key in \texttt{Elliptic} design.  As we increase the control steps of a design to meet latency requirement, we see a decreasing correlation coefficient  $(r_{1}^{2})$ resulting in higher MTD$_{1}$. As more number of bits ($p$) switch, the value of MTD$_{0}$ decreases. However, it would also increase the switching power excessively and may damage the IP core before an attacker can retrieve the key. We report the key extraction probability in a design lockout architecture in Table  \ref{tab:key_prob}. For finite control steps and larger key size, DLockout provides a negligible probability that a key bit can be leaked and it is independent of any RTL obfuscation technique and key size. Across all benchmarks, we see an average area overhead in Fig. \ref{fig:singleplot} increase from 1.50\% to 7.78\%. The average delay overhead ranges in between 0.1\% to 9.59\%. With the increase in key length, the number of nets increases as well in Fig. \ref{fig:singleplot} (d) which makes  the probability of finding the nets responsible for key propagation small.

\begin{table}
\begin{center}
\caption{Number of traces to retrieve a single key in DPA attack for key size(=32bit)}
\vspace{-1ex}
\label{tab:dpa_meas}
{\begin{tabular}{|c|c|c|c|c|c|c|}
\midrule
$M$ & $N$ & $q$ & $p$ & MTD$_{0}$ & r$_{1}^{2}$ & MTD$_{1}$ \\ \midrule
\multirow{9}{*}{32} & \multirow{3}{*} {4} & \multirow{3}{*} {32} & 8  & 4  &  0.060 & 8533\\
& & & 16 & 2 & 0.028 & 9142\\
& & & 32 & 1 & 0.011 & 12800\\ \cline{2-7}
& \multirow{3}{*} {5} & \multirow{3}{*} {32} & 8 & 4   & 0.055 & 11636\\ 
& & & 16 & 2 & 0.022 & 14545\\ 
& & & 32 & 1 & 0.009 & 17777\\ \cline{2-7}
& \multirow{3}{*} {6} & \multirow{3}{*} {32} & 8 & 4  & 0.051 & 15058\\ 
& & & 16 & 2  & 0.020 & 19200\\ 
& & & 32 & 1 & 0.007 & 27428\\ \cline{1-7}
\end{tabular}}
\begin{tablenotes}
 \item[1] $M$: key size; $N$: number of control steps; $q$: bit width of a MUX input; $p$: number of bits that switch in a MUX input; $C$ = 1.
\end{tablenotes}
\end{center}
\vspace{-6ex}
\end{table}



\begin{table}
\begin{center}
\caption{Key extraction probability for three key length and five attempts}
\vspace{-1ex}
\label{tab:key_prob}
{\begin{tabular}{|c|c|c|c|c|}
\midrule
($m$,$n$) & $P(m,n)$ & $X$ & $K$ & $f(K,X,P(m,n))$\\ \midrule
\multirow{5}{*}{(32,32)} & \multirow{5}{*} {0.08e-44} & \multirow{5}{*} {5} & 1 & 0.4e-44\\
& & & 2 & 0.8e-44\\
& & & 3 & 0.8e-44\\
& & & 4 & 0.4e-44\\ 
& & & 5 & 0.08e-44\\ \midrule
\multirow{5}{*}{(64,64)} & \multirow{5}{*} {0.43e-108} & \multirow{5}{*} {5} & 1 & 2.15e-108\\
& & & 2& 4.3e-108\\
& & & 3& 4.3e-108\\
& & & 4& 2.15e-108\\ 
& & & 5& 0.43e-108\\ \midrule
\multirow{5}{*}{(128,128)} & \multirow{5}{*} {0.07e-253} & \multirow{5}{*} {5} & 1& 0.35e-253\\
& & & 2& 0.7e-253\\
& & & 3& 0.7e-253\\
& & & 4& 0.35e-253\\ 
& & & 5& 0.07e-253\\ \midrule
\end{tabular}}
\begin{tablenotes}
 \item[1] $m$: key size; $n$: number of obfuscation logic; $X$: Allowed attempts; $K$: Attempt index
\end{tablenotes}
\end{center}
\vspace{-6ex}
\end{table}

\section{Conclusions}
\label{sec:conclusion}
In this paper, we propose DLockout, that can provide an add-on to the existing obfuscated RTL IP to increase the difficulty in reverse engineering. Once the number of key recovery attempts exceeds the preset threshold, the design is self-locked out to provide strong security guarantee against brute force attacks. In future, we plan to extend the work to provide technique for genuine user to recover after circuit locking.  The effects of DLockout architecture on design parameters are minimal and modifications on DLockout architecture are  presented against side channel attacks.





%

\bibliographystyle{unsrt}
\scriptsize{
\bibliography{bib/HT,bib/RTL,bib/metering,bib/watermark,bib/miscell,bib/fingerprint,bib/camof,bib/split_manu,bib/logic_encryp,bib/logic,bib/pa}
}


\end{document}